\documentstyle[12pt,amsbsy]{article}
\newcommand{\beq}{\begin{equation}}
\newcommand{\eeq}{\end{equation}}

\newcommand{\beqn}{\begin{eqnarray}}
\newcommand{\eeqn}{\end{eqnarray}}

\newcommand{\p}{\mbox{${\bf p}$}}
\newcommand{\q}{\mbox{${\bf q}$}}

\newcommand{\bv}{\mbox{${\bf v}$}}
\newcommand{\s}{\mbox{${\bf s}$}}

\newcommand{\si}{\mbox{${\boldsymbol\sigma}$}}
\newcommand{\Si}{\mbox{${\boldsymbol\Sigma}$}}

\newcommand{\E}{\mbox{${\bf E}$}}
\newcommand{\B}{\mbox{${\bf B}$}}

\newcommand{\fib}{\mbox{${\boldsymbol\phi}$}}

\newcommand{\ep}{\mbox{${\epsilon}$}}

\begin{document}

\begin{titlepage}

\vspace{1cm}

\begin{center}
{\large \bf Quadrupole interaction of relativistic quantum particle\\
with external fields}
\end{center}

\begin{center}
A.A. Pomeransky\footnote{pomeransky@vxinpz.inp.nsk.su}, 
R.A. Sen'kov\footnote{senkov@vxinpz.inp.nsk.su}
\end{center}
\begin{center}
Budker Institute of Nuclear Physics\\
630090 Novosibirsk, Russia
\end{center}

\bigskip

\begin{abstract}
We consider the motion of a spinning relativistic particle with
an arbitrary value of spin in external electromagnetic and
gravitational fields, to first order in the external field.
We use the noncovariant description of spin. An explicit
expression is obtained for the interaction of second order in spin.
The value of the quadrupole moment is found for which this
interaction decreases when the energy grows.
\end{abstract}

\end{titlepage}

\section{Introduction}

The problem of motion of a particle with internal angular momentum
(spin) in external fields has been considered in many papers
(see \cite{kp} and references therein).
The problem has various applications, ranging from the accelerator
physics to the dynamics of rotating stars or black holes.
Our interest to the problem originated from the question about
evolution of polarisation of nuclei in storage rings.

In the present paper we discuss in detail quadratic in spin
effects. We consider only the terms proportional
to first derivatives of the field strength, neglecting higher
derivatives.
Thus electric quadrupole interaction for arbitrary particle
velocities
will be found. The linear in spin interaction (magnetic dipole one)
is well-known (see, e.g., \cite{blp}, \S 41), so we will not write it
down here.

We find the interaction of a spinning relativistic particle
with external gravitational field to the same approximations.

We use here the approach
described in Refs. \cite{kp}, \cite{kms}. It allows one to
obtain equations of motion of a particle in external fields,
to an arbitrary order in spin. However we confine neither to
semiclassical approximation, in contrast to \cite{kp},
nor nonrelativistic approximation, in contrast to \cite{kms}.

\bigskip

\section{Spinning particle in electromagnetic field}

The Lagrangian of the spin interaction with an external
field can be derived from the elastic scattering amplitude
\beq\label{ampl}
- e J^{\mu}A_{\mu}
\eeq
of a particle with spin $s$ on a vector potential
$A_{\mu}$ \cite{kp}.
The matrix element $J_{\mu}$ of the electromagnetic
current operator between states with momenta $k$ and
$k^\prime$ can be written (under $P$ and $T$ invariance)
as follows (see [1-3]):
\beq\label{cur}
J_{\mu}=\frac{1}{\sqrt{\ep_k\ep_{k^\prime}}}\,\bar{\psi}(k^\prime)
\left\{p_{\mu}F_e +\, \frac{1}{2}\Sigma_{\mu\nu}q^{\nu}\,F_m\right\}
\,\psi(k).
\eeq
Here $\;p_{\mu}=(k^\prime+k)_{\mu}/2,\;\;
q_{\mu}=(k^\prime-k)_{\mu}$.
The wave function of a particle with an arbitrary spin
$\psi$ can be written (see, for instance, Ref.
\cite{blp}, \S 31) as
\begin{equation}\label{ps}
            \psi={1 \over \sqrt 2} \left( \begin{array}{c} \xi\\
                                         \eta\\
                        \end{array}
                 \right).
\end{equation}
Both spinors,
\[ \xi=\{ \xi^{{\alpha}_1\,{\alpha}_2\,\,..\,\,
{\alpha}_p\,}_{\dot{\beta}_1
\,\dot{\beta}_2\,..\,\,\dot{\beta}_q}\} \]
and
\[ \eta=\{ \eta_{\dot{\alpha}_1\,\dot{\alpha}_2
\,\,..\,\,\dot{\alpha}_p\,}^{{\beta}_1\,{\beta}_2\,..\,\,
{\beta}_q}\},
\]
are symmetric in the dotted and undotted indices separately,
and
\[ p+q=2s. \]
For a particle of a half-integer spin one can choose
\[ p=s+\,\frac{1}{2}\,,\;\;\;\;q=s-\,\frac{1}{2}\,\,. \]
In the case of an integer spin it is convenient to put
\[ p=q=s. \]
The spinors $\,\xi\,$ and $\,\eta\,$ are chosen in such a way
that under reflection they go over into each other (up to a
phase). At $p \neq q$ they are different objects which
belong to different representations of the Lorentz group.
If $p=q$, these two spinors coincide. Nevertheless, we will
use the same expression (\ref{ps}) for the wave function of
any spin, i.e., we will also introduce formally the object
$\,\eta\,$ for an integer spin, keeping in mind that it is
expressed in terms of $\,\xi\,$. This will allow us to
perform calculations in the same way for the integer and
half-integer spins.

In the rest frame both $\xi$ and $\eta$ coincide with a
nonrelativistic spinor $\xi_0$, which is symmetric in all
indices; in this frame there is no difference between dotted
and undotted indices. The spinors $\xi$ and $\eta$ are
obtained from $\xi_0$ through the Lorentz transformation:
\beq\label{lt}
\xi=\exp\{\Si\fib/2\} \xi_0\,;\;\;\;\;\;\;\;
\eta=\exp\{-\Si\fib/2\} \xi_0\,.
\eeq
Here the vector $\fib$ is directed along the velocity,
$\;\;\tanh\phi=v$;
\[  \Si \,=\,\sum_{i=1}^{p} \si_i\,-\,
\sum_{i=p+1}^{p+q} \si_i\,, \]
and $\si_i$ acts on the $i$th index of the spinor $\,\xi_0\,$
as follows:
\begin{equation}\label{aa}
\si_i\,\xi_0
=(\si_i)_{\alpha_i\beta_i}\, (\xi_0)_{....\beta_i...}\;.
\end{equation}
In the Lorentz transformation (\ref{lt}) for $\xi$, after
the action of the operator $\Si$ on $\xi_0$ the first $p$
indices are identified with the upper undotted indices and
the next $q$ indices are identified with the lower dotted
indices. The inverse situation takes place for $\eta$.

Then,
\[ \bar{\psi} = \psi^\dagger \gamma_0 =
             \psi^\dagger  \left(
                            \begin{array}{rr}
                                0 & I \\
                                I & 0\\
                             \end{array}
                       \right);
      \]
here $I$ is the sum of unit $2\times 2$ matrices acting on
all indices of the spinors $\,\xi\,$ and $\,\eta\,$. The
components of the matrix $\Sigma_{\mu\nu}=-\Sigma_{\nu\mu}$
are:
\begin{equation}\label{0n}
      \Sigma_{0n}= \left(
                            \begin{array}{rr}
                             -\Sigma_n    & 0 \\
                                 0 & \Sigma_n\\
                             \end{array}
                       \right);
      \end{equation}
\begin{equation}\label{mn}
      \Sigma_{mn}=\,-\,2i\ep_{mnk}\left(
                            \begin{array}{rr}
                                s_k & 0  \\
                                0   & s_k\\
                             \end{array}
                       \right);
      \end{equation}
\[ \s =\,\frac{1}{2}\sum_{i=1}^{2s} \si_i.  \]

The scalar operators $F_{e,m}$ depend on two invariants,
$t=q^2$ and 
$\tau=(Sq)^2$, here $S^\mu=(i/4)\epsilon^{\mu\nu\kappa\lambda}
\Sigma_{\kappa\lambda}p_{\nu}/m$ is the spin 4-vector.
In the expansion in the electric multipoles
\[ F_e(t,\tau)=\sum_{n=0}^{N_e}f_{e,2n}(t)\tau^n \]
the highest power $N_e$ equals obviously to $s$ and
$s-1/2$ for an integer and half-integer spin, respectively.
In the magnetic multipole expansion
\[ F_ m(t,\tau)=\sum_{n=0}^{N_m}f_{m,2n}(t)\tau^n \]
the highest power $N_m$ constitutes $s-1$ and $s-1/2$ for an
integer and half-integer spin. It can be easily seen that
$$f_{e,0}=1,\;\;\;\;f_{m,0}=\frac{g}{2}.$$
Here and below we confine to the consideration of formfactors
at zero momentum transfer, $f_{e,0}=f_{e,0}(0), f_{m,0}=f_{m,0}(0).$

In our consideration we will discuss terms linear and quadratic in
the momentum transfer momentum ${\bf q}$ only, i.e., proportional
to the field strength and its derivatives. We will write down
quadratic in spin terms only, because linear in spin interaction
is well-known (see, for instance, \cite{blp}, \S 41). 

We will start with the contribution of the convection term
\beq\label{conv}
-\,\frac{e}{\sqrt{\ep_k\ep_{k^\prime}}}\,\bar{\psi}(k^\prime)\psi(k)
\,p^{\mu}A_{\mu}.
\eeq
The product of exponents in the expression
\beq\label{pro}
\bar{\psi}(k^\prime)\psi(k)= {1 \over 2}\xi_0^{\prime \dagger}
[\exp\{\Si\fib^\prime/2\} \exp\{-\Si\fib/2\}
+\exp\{-\Si\fib^\prime/2\}\exp\{\Si\fib/2\}]\xi_0
\eeq
can be calculated analogously to \cite{kp}.
To our accuracy it equals to
\beq\label{prod}
\exp\{\Si\fib^\prime/2\} \exp\{-\Si\fib/2\}
=\exp\left\{\frac{1}{2m}\Si\q-\frac{\Si\p(\p\q)}{2m\ep(\ep+m)}+
i\frac{\s[\p \times \q]}{m(\ep+m)}\right\},
\eeq
where $\ep=\sqrt{m^2+\p^2}=m\gamma $.
Thus wave functions scalar product takes the form
\beq\label{produ}
\bar{\psi}(k^\prime)\psi(k)= 
\xi_0^{\prime \dagger}\left\{1+i\frac{\s[\p \times \q]}{m(\ep+m)}-
\frac{(\s[\p \times \q])^2}{2m^2(\ep+m)^2}+
\frac{1}{8m^2}\left(\Si\q-\frac{\Si\p(\p\q)}{\ep(\ep+m)}\right)^2+
O(q^3)\right\}\xi_0.
\eeq
The contribution to the Lagrangian of the term from (\ref{produ}),
which contains $\Si$ and therefore vanishes in the classical limit,
is
\beq\label{qu1}
e\,\frac{\Lambda}{2m^2}\,\left[(\s\nabla)\,-\,
{\gamma \over \gamma+1}\, (\bv\s)(\bv \nabla)\right]\,\left[(\s\E)\,
-\,{\gamma \over \gamma+1}\,(\s\bv)(\bv\E)\,
+\,(\s[\bv\times\B])\right],
\eeq
where
\begin{eqnarray}\label{Lam}
 \Lambda=\left\{
                   \begin{array}{cc}
                              1/(2s-1), & \mbox{integer spin,}\\
                              1/(2s), & \mbox{half-integer spin.}\\
                   \end{array}
             \right. 
\end{eqnarray}
We consider the motion of particles in empty space (for example,
in an accelerator), so in the derivation of (\ref{qu1}) we used
sourceless
Maxwell equations. Here and below we use the orthogonality condition
$pq\,=\,0$, which means that initial and final particles are on
mass-shell. In the nonrelativistic approximation the contribution
(\ref{Lam}) has been derived in \cite{kms} (including the additional
so-called Darwin term corresponding to the contact interaction with
external sources).

The remaining part of the convection contribution to the
Lagrangian is
\beq\label{lse1}
-\,{e \over 2m^2}\,{\gamma \over \gamma+1}\,\left(\s\,[\bv\times
\nabla]\right)\left[\left(1-\,{1 \over \gamma}\right)(\s\B)\,-\,
{\gamma \over \gamma+1}\,(\s\bv)(\bv\B)\,-\,
{\gamma \over \gamma+1}\left(\s\,[\bv\times \E]\right)\right].
\eeq

This expression which has nonvanishing classical limit was derived
in \cite{kp}.

Let us go over to the terms proportional to $g$-factor. Their
classical part also was found in \cite{kp}. The part vanishing
in the classical limit is
\beq\label{qu2}
-\,e\,g\,\frac{\Lambda}{2m^2}\,\left[(\s\nabla)\,-\,
{\gamma \over \gamma+1}\, (\bv\s)(\bv \nabla)\right]\,\left[(\s\E)\,
-\,{\gamma \over \gamma+1}\,(\s\bv)(\bv\E)\,
+\,(\s[\bv\times\B])\right].
\eeq

At last, the "bare" quadrupole interaction, which enters the
expressions
(\ref{ampl}) and (\ref{cur}), is
\beq\label{qu}
-\,\frac{e}{\sqrt{\ep_k\ep_{k^\prime}}}\,f_{e,2}\tau
p^{\mu}\,A_{\mu}.
\eeq
The corresponding contribution of the "bare" quadrupole interaction
to the Lagrangian is
\beq\label{ququ}
-\,e\,f_{e,2}\,\left[(\s\nabla)\,-\,
{\gamma \over \gamma+1}\, (\bv\s)(\bv \nabla)\right]\,\left[(\s\E)\,
-\,{\gamma \over \gamma+1}\,(\s\bv)(\bv\E)\,
+\,(\s[\bv\times\B])\right].
\eeq
The general expression for the quadratic in spin interaction
can be presented as follows
\[
L_{2s}=\,-\,e\,\left(f_{e,2}+\frac{\Lambda(g-1)}{2m^2}\right)\,
\left[(\s\nabla)\,-\,
{\gamma \over \gamma+1}\, (\bv\s)(\bv \nabla)\right]\,\left[(\s\E)\,
-\,{\gamma \over \gamma+1}\,(\s\bv)(\bv\E)\,
+\,(\s[\bv\times\B])\right]
\]
\[
+\,{e \over 2m^2}\,{\gamma \over \gamma+1}
\,\left(\s\,[\bv\times \nabla]\right)
\left[\left(g-1+\,{1 \over \gamma}\right)(\s\B)\,-\,
(g-1)\,{\gamma \over \gamma+1}\,(\s\bv)(\bv\B)\,\right.
\]
\beq\label{total} 
\left. -\,\left(g-{\gamma \over \gamma+1}\right)
\left(\s\,[\bv\times \E]\right)\right].
\eeq
The obtained expression for the quadratic in spin interaction
differs from the corresponding semiclassical formula in \cite{kp}
by the following substitution only
\beq\label{zam}
f_{e,2} \rightarrow f_{e,2}+\frac{\Lambda(g-1)}{2m^2}
\eeq
There is a value of $f_{e,2}$, for which the quadratic in spin
interaction $L_{2s}$ decreases for $\;\gamma\rightarrow \infty\,$.
An analogous phenomenon exists for the linear in spin interaction:
for $g = 2$ it decreases when the particle energy grows.
The corresponding preferred value of the formfactor $f_{e,2}$
can be easily obtained by substituting (\ref{zam}) into the
corresponding expression in \cite{kp}:
\beq\label{ff}
f_{e,2} = (1-\Lambda)\frac{g-1}{2m^2};\;\;\;\;\;\;{\rm for}\;\;\;
g=2\;\;
f_{e,2}=\frac{1-\Lambda}{2m^2}.
\eeq
Let us write down the preferred value of the quadrupole moment:
\beq\label{qua}
Q\,=\,Q_{zz}\vert_{s_z=s}=\,-\,2\,e\,
\left(f_{e,2}+\frac{\Lambda(g-1)}{2m^2}\right)\,s(2s-1)=\,-\,
\frac{e\,s(2s-1)}{m^2}.
\eeq
The same preferred value of the quadrupole moment was recently
derived
in \cite{susy} from different approach -- based on the
supersymmetric
sum rules.
This good high-energy behavior of the interaction
is a necessary (but insufficient) condition of
renormalizability. Indeed, charged vector boson in the
renormalizable electroweak theory has the quadrupole moment
equal to
\beq
Q\,=\,-\frac{e}{m^2},
\eeq
in accordance with the expression (\ref{qua}).

\bigskip

\section{Spinning particle in gravitational field}

The equations of motion in an external gravitational field to any
order in spin can be obtained from the equations of
motion in an electromagnetic field by simple substitution.

The elastic scattering amplitude in a weak
external gravitational field $h_{\mu\nu}$ is
\beq\label{gampl}
-\,{1 \over 2}T_{\mu\nu}h^{\mu\nu}
\eeq
(in due time we will go over to a generally covariant form).
The matrix element $T_{\mu\nu}$ of the energy-momentum tensor
between the states of momenta $k$ and $k^\prime$ can be written
as \cite{kp}:
\[ T_{\mu\nu}=\frac{1}{4\;\sqrt{\ep_k\ep_{k^\prime}}}\,\bar{\psi}
(k^\prime)
\left\{4\;p_{\mu}p_{\nu}\,F_1
+\,(p_{\mu}\Sigma_{\nu\lambda}
+\,p_{\nu}\Sigma_{\mu\lambda})\,q^{\lambda}\,F_2 \right. \]
\beq\label{tens}
\left. +\,(\eta_{\mu\nu}q^2\,-\,q_\mu q_\nu)\,F_3
+\,[S_\mu S_\nu q^2 -\,(S_\mu q_\nu+\,S_\nu q_\mu)(Sq)\,+\,
\eta_{\mu\nu}(Sq)^2]\,F_4 \right\}\,\psi(k).
\eeq
The scalar operators $F_i$ in this expression are also expanded
in powers of $\tau=(Sq)^2$:
\beq\label{expan}
F_i(t,\tau)=\sum_{n=0}^{N_i}f_{i,2n}(t)\tau^n.
\eeq
Since we are interested in the equations of motion in a
sourceless field, the terms proportional to $F_3$ and $F_4$ in
the expansion (\ref{tens}) will be omitted, because when rewritten
in
the covariant form, they are proportional to the scalar curvature
and Ricci tensor, respectively. So, the amplitude (\ref{gampl})
can be
presented in the following form
\beq\label{gampl2}
-\,\frac{1}{2\,\sqrt{\ep_k\ep_{k^\prime}}}\,\bar{\psi}(k^\prime)
\left\{p_{\mu}\,F_1+\,{1 \over 2}\Sigma_{\mu\lambda}\,
q^{\lambda}\,F_2
\right\}\,\psi(k)h^{\mu\nu}\,p_{\nu}.
\eeq
Clearly, (\ref{gampl2}) differs (\ref{ampl}), (\ref{cur})
by the following substitution only:
\beq\label{zam2}
e\,A_{\mu} \rightarrow {1 \over 2}\,h_{\mu\nu}\,p^{\nu}.
\eeq
With this substitution in (\ref{total}), one can obtain quadratic
in spin interaction of a particle with a gravitational field.

Let us now consider the preferred values of the
gravitational formfactors. The first coefficients in the expansion
(\ref{expan}) $f_{1,0}=1$ and $f_{2,0}=1$ are fixed by the general
covariance [6,7], and they coincide with the corresponding
electromagnetic
formfactors at zero momentum transfer if $g=2$. Due to analogous
structure of the amplitudes (see (\ref{zam2}))
and to the coincidence of the first coefficients in the formfactors
expansion,
the preferred values of formfactors coincide for electromagnetic and
gravitational interactions:
$f_{e,2n}=f_{1,2n}$ and $f_{m,2n}=f_{2,2n}$.
In particular, it follows from (\ref{ff}) that
\beq
f_{1,2}=\frac{1-\Lambda}{2m^2}.
\eeq
Let us note that the Lagrangian corresponding to the
interaction with $f_{1,2}$, can be rewritten in the generally
covariant form \cite{kp}, \cite{kh}:
\begin{equation}\label{gm}
{L}_{gm}=\frac{\kappa}{8m}R_{abcd}S^{ab}S^{cd}\,.
\end{equation}
Here the dimensionless parameter $\kappa$
(the gravimagnetic ratio \cite{kp}) is related to
$f_{1,2}$ as follows:
\beq
\kappa=\frac{2m^2\,f_{1,2}}{1-\Lambda}.
\eeq
So, the preferred value of the parameter $\kappa$ is
\beq
\kappa=1.
\eeq
The preferred value of the gravimagnetic ratio
$\kappa=1$ was also derived in [1,8,9] and,
on the basis of the supersymmetric sum rules, in \cite{sus2}.

\bigskip

We are grateful to I.B. Khriplovich who brought this problem
to our attention for many useful discussions and critical
reading of the manuscript.
The work was supported by the Russian
Foundation for Basic Research through Grant No. 98-02-17797 and
Grant No. 96-15-96317 for the support of leading scientific schools
of RF;
Ministry of Education, Grant No. 3H-224-98.

\end{document}